\documentclass[a4paper,11pt]{article}
\usepackage[utf8]{inputenc}
\usepackage[margin=2cm]{geometry}
\usepackage{multicol}
\usepackage{amsmath,amssymb,isotope}
\usepackage{graphicx}
\usepackage{caption}
\usepackage{subcaption}
\usepackage{color}
\usepackage[backend=biber,style=ieee]{biblatex}
\usepackage[title]{appendix}
\newcommand{\be}{\begin{eqnarray}}
\newcommand{\ee}{\end{eqnarray}}
\newcommand{\nn}{\nonumber}

\title{Enhanced plateau effect at resonance \\in realistic non-integrable EMRIs}
\author{Areti Eleni \thanks{email: aeleni@phys.uoa.gr} \\
\footnotesize{Section of Astrophysics, Astronomy, and Mechanics, Department of Physics,} \\
\footnotesize{University of Athens, Panepistimiopolis Zografos GR15783, Athens, Greece} \and 
Theocharis A. Apostolatos \thanks{email: thapostol@phys.uoa.gr}\\
\footnotesize{Section of Astrophysics, Astronomy, and Mechanics, Department of Physics,} \\
\footnotesize{University of Athens, Panepistimiopolis Zografos GR15783, Athens, Greece}}

\begin{document}

\maketitle
\begin{abstract}
When an EMRI in a perturbed integrable gravitational field,
such as a deformed Kerr black hole,
undergoes a prolonged  resonance, the frequencies that engage in resonance retain a fixed rational ratio, despite experiencing adiabatic changes due to radiation reaction. In the past this plateau effect in the evolution of the ratio of frequencies has been investigated by studying the
orbital evolution through
kludge models, which provide approximate average losses of energy and angular momentum experienced by a test particle in this field.
By employing a Newtonian gravitational field
that closely resembles a pure Kerr or 
a perturbed Kerr relativistic field, we demonstrate that the 
actual adiabatic evolution of an orbit driven by an artificial ``self-force'' results in more prolonged periods of resonance crossings compared to those obtained by imposing a predetermined  rate of energy and angular momentum change throughout the orbital progression..
\end{abstract}


\section{Introduction}
Extreme mass ratio inpirals (EMRIs), are prominent sources of gravitational waves (GWs) for the future space-based detector Laser Interferometer Space Antenna (LISA) \cite{LISA}. EMRIs are  binaries consisting of a stellar mass compact object, i.e., a black hole (BH) or a neutron star (NS) of  mass $m$,  inspiraling around a supermassive BH of  mass $M$, with mass ratio $\epsilon=m/M\leq 10^{-4}$.

Since the lighter compact object of an EMRI spends the last few years of its inspiral tracing out the strong-gravity  region of the supermassive BH, EMRIs offer us the opportunity to test the theory of General Relativity (GR) and its astrophysical implications concerning the formation of black holes.
The last hundreds of thousands of GWs cycles 
of such a system encode the details of the spacetime geometry
of the massive object; thus by analysing these waves, one could read out its multipole moments \cite{Ryan}.

According to GR the gravitational field of an astrophysical BH is described by the Kerr metric \cite{Kerr}, the multipole moments  of which are determined only by its mass and spin \cite{Geroch,Hanse}. 
Since the Kerr metric is characterized by a few symmetries, the equations governing the geodesic motion around a Kerr BH form an integrable system.
The conservation of the energy and $z$-angular momentum along the axis of symmetry is associated with a time-translation and a rotational Killing vector, respectively, while the existence of the Carter constant \cite{Carter} is linked to a hidden symmetry of a rank-two Killing tensor. 
As a consequence, a bound geodesic Kerr orbit in the spatial part of the phase-space is confined to lie on a compact torus characterized by three fundamental frequencies \cite{Schmidt}. 
Trajectories ergodically fill the phase-space  tori, unless two or more fundamental frequencies form a rational ratio (resonant orbits).

However, the actual orbit of the small compact object around the massive BH is not exactly geodesic, due to the gravitational self-force (SF) which arises from the object's interaction with the time-dependent gravitational field \cite{Mino,Quinn}. 
The dissipative part of SF drives the object to a gradual inspiral towards the massive BH, following an adiabatic evolution of geodesics,  while it radiates away energy and angular momentum in the form of GWs.
The orbital motion is obtained from BH perturbation theory with the small mass ratio $\epsilon$ as an expansion parameter \cite{Poisson, Barack}.
The SF for a non-spinning particle on a generic orbit around a Kerr BH to first order in the mass ratio $\epsilon$ has been obtained recently \cite{Van}.

During the inpiral the three orbital fundamental frequencies change slowly, thus a resonance will occur when
two of them form a rational ratio.
Usually, the methods used for computing the orbital evolution, and the corresponding waveforms, become inadequate at resonance, where the ``constants" of motion then change rapidly, leading to large shifts in waveform's phase \cite{FlanaHinder}. 

An important characteristic of resonances is that they can be used to discriminate if the background spacetime is not an integrable Kerr one, because either the central BH is not described by GR  or the environment of the BH is not vacuum.
Such spacetimes probably don't possess all the special symmetries of Kerr that lead to a third integral of motion and form a complete integrable system. 
These cases could be described  as appropriate deformations of Kerr metric.
However, when an integrable Hamiltonian is slightly perturbed, its phase-space tori undergo changes.
The Poincar{\'e}-Birkhoff theorem \cite{PoincBirk1, PoincBirk2} states that the resonant tori desintegrate and form islands of stability (Birkhoff islands), occupying a phase-space volume of non-zero measure, inside which the ratio of frequencies remain locked to a constant rational value.
Birkhoff islands are characteristic features of non-integrable dynamical systems.

Ref. \cite{AposGeraCont} investigated the evolution of the ratio of the orbital frequencies of a particle orbiting around a non-Kerr object describing by the Manko-Novikov (MN) metric \cite{Manko}, when its trajectory crosses a Birkhoff island.
Due to the lack of an expression for the radiation reaction SF for  non-Kerr spacetimes, the numerical integration of an orbit was performed by combining the equations of geodesic motion for MN metric with the hybrid approximative method \cite{hybrid} which provides the average losses of energy and $z$-angular momentum.
Assuming constant rates of energy and $z$-angular momentum losses,
the time interval within which the orbit remains at a prolonged resonance (i.e., stays in a Birkhoff island) was computed.
During that time both frequencies change while their ratio remains constant.
Whenever such a plateau in the evolution of the ratio of frequencies is observed one could conclude that the central object is not a Kerr BH. 
Also, in \cite{destounis}, following  a similar procedure, Destounis et al. found that when the orbit crosses a prolonged resonance the GW frequency appears a rapid but short-lived  "glitch".

In the present work we would like to address the question  whether the assumption of constant rates of change of energy and $z$-angular momentum leads to  wrong estimates of the time interval of resonance crossings.
Lacking a SF formula for a non-Kerr spacetime, we will resort to a Newtonian analogue problem.

In Ref. \cite{Eleni} it has been shown that  the Euler gravitational field of a pair of spatially-fixed point masses at an imaginary distance from each other, is a very good analogue of the Kerr relativistic field. 
Moreover this particular field can be modified so as to transform the system from an integrable to a slightly non-integrable one. 
By incorporating an additional small external dissipative force, we could drive adiabatically an orbit,
in a similar fashion that a geodesic orbit is driven in a given background spacetime by the radiation reaction caused by a self-force.
At the same time, the average losses of energy and $z$-angular momentum in the adiabatic limit for such a dissipative force are computed.
Once again, the orbit is evolved by a new integration scheme, based on inducing  the subsequent time depending ``constants'' of motion, but without any direct dissipative force implied. Finally the two distinctive 
numerical schemes were compared with respect to the total resonance crossing time. There was a systematic 
enhancement of the crossing time, by a factor of at least 2, when the  instantaneous dissipative force was employed.


The rest of the article is organized as follows: In Section 2 an overall description of the oblate Euler problem is given. In Section 3 we describe the perturbed version of this problem, constructed by introducing a small mass at the midpoint between the two fixed masses. In Section 4 we give a brief description of some theoretical features of slightly non-integrable problems. In Section 5, we 
introduce the dissipative force that is used and explain the two different integration schemes 
followed to drive an orbit in the perturbed Euler field.
The scheme based on average losses is further analysed in Section 6. Finally, in Section 7 we present our results
and discuss their implications.



\section{The oblate Euler problem}

The Euler problem of two fixed centers 
\cite{Euler} describes the gravitational field of two static point masses $m_1$ and $m_2$ at a fixed distance $2a$ between them.
We assume that the $z-$axis is the axis along which the two masses are located at $z_1= a \hat{z}$ and $z_2=- a \hat{z}$, respectively, with $a$ being constant and real.
By setting the two masses equal to each other, i.e., $m_1=m_2=M/2$, and their distance imaginary, i.e., $a\to ia$, the potential becomes oblate (with negative quadrupole moment) and can be considered as the Newtonian analogue of the relativistic Kerr black hole \cite{Kerr,Keres,Israel,Eleni,Will}.
We need the symmetric case with equal masses because only then 
the gravitational potential of each mass is the complex conjugate to the potential of the other mass, allowing the combined potential field of the two masses 
to be real.
The resulting gravitational field of the oblate Euler (also known as Vinti potential in astronomy, used to describe the gravitational field around oblate planets \cite{Vint}), is stationary, axisymmetric along the $z-$axis, and reflection symmetric along the equatorial plane described by the following form:
\begin{equation}\label{V0}
V_0=
-\frac{G (M/2)}{|{\bf r}- i a {\bf \hat{z}}|}
-\frac{G (M/2)}{|{\bf r}+ i a {\bf \hat{z}}|},
\end{equation}
where ${\bf r}$ is the radial distance from the axes origin and by $|{\bf k}|$ we mean $\sqrt{{\bf k}\cdot{\bf k}}$.
The latter vector product is  a complex number and in order to keep the square root single-valued we should adopt a branch cut. 
We have chosen the negative real axis of the vector product as the branch-cut of our potential function so that the two denominators in \eqref{V0} will be conjugate to each other leading to a real potential. Henceforth, when we mention the Euler field, we shall exclusively refer to the oblate Euler field, and later on to its perturbed version.

A general stationary, axisymmetric and reflection symmetric along the equatorial plane Newtonian potential that vanishes at infinity can be fully decomposed in multipole moments $M_l$  through the relation \cite{Will,Glampedakis}:

\begin{equation*}
    V=-\sum_{l=0}^{\infty} \frac{M_{2l}}{{r}^{2l+1}}P_{2l}(z/r),
\end{equation*}
where $P_l$ are the Legendre polynomials.
It turns out that the  multipole moments of the  Euler potential (\ref{V0}) is given by \cite{Eleni,Glampedakis}:

\begin{equation*}
    M_{2l}=M(-a^2)^l,
\end{equation*}
which is the same as the ``no-hair'' relation obeyed by the mass multipole moments of the Kerr metric \cite{Geroch,Hanse} with the length parameter $a$ of the Euler field, playing the role of the spin of a Kerr black hole \cite{Eleni,Will}.

A more appropriate coordinate system to study the motion in this  field is that of oblate spheroidal coordinates, $(\xi, \eta, \phi)$, 
where $\phi$ is the usual spherical azimuthal angle, $\xi \in [0,+\infty)$ and  $\eta \in [-1,1]$. 
These new coordinates are related to the Cartesian coordinates $(x,y,z)$ by:
\begin{align*}
x &=a \sqrt{(1+\xi^2)(1-\eta^2)}\cos{\phi},  \\
y &=a \sqrt{(1+\xi^2)(1-\eta^2)}\sin{\phi},  \\
z &=a \xi \eta,
\end{align*}
and to the spherical coordinates $(r,\theta)$ by:
\begin{align*}
    r&=a\sqrt{1+\xi^2-\eta^2},\\
    \cos{\theta}&=\frac{\xi\eta}{\sqrt{1+\xi^2-\eta^2}}.
\end{align*}

In terms of oblate spheroidal coordinates the Euler potential
\eqref{V0} assumes the following simple form:
\begin{align}\label{VoE}
V_0(\xi,\eta)=-
\frac{G M_0 \xi }{a (\xi^2+\eta^2)}.
\end{align} 
From the above multipole expansion, $M_0=M$.
It should be noted that the field $V_0(\xi, \eta)$ is defined everywhere except when $\xi=0$ and $\eta=0$, which corresponds to the equatorial focal circle $(r=a, \theta=\pi/2)$, where the potential becomes singular. This singularity corresponds to Kerr's ring singularity.

The motion of a test particle in  the Newtonian  Euler potential is independent of it's mass, so the Hamiltonian (per unit test-particle mass $\mu$) is:

\begin{equation}\label{Ham}
   H_0=\frac{1}{2a^2} \left[
p_\xi^2 \frac{\xi^2+1}{\xi^2+\eta^2} + 
p_\eta^2 \frac{1-\eta^2}{\xi^2+\eta^2}+\frac{p_\phi^2}{(\xi^2+1)(1-\eta^2)}\right] +V_0(\xi,\eta). 
\end{equation}

The conjugate momenta to $\xi,\eta,\phi$ are defined as:
\begin{align}
p_\xi&= a^2 \frac{\xi^2+\eta^2}{\xi^2+1} \dot{\xi},\label{pxi}\\
p_\eta&=a^2 \frac{\xi^2+\eta^2}{1-\eta^2} \dot{\eta},\label{peta}\\
p_\phi&= a^2 (\xi^2+1)(1-\eta^2) \dot{\phi}\label{pphi}, 
\end{align}
where ``$~\dot{}~$'' denotes time derivative.

The stationarity and axisymmetry of the system is obvious in the Hamiltonian expression $H_0$.
The time and azimuthal coordinate are cyclic, leading to conservation of the energy $E=H_0$ and the angular momentum along the axis of symmetry $L_z=p_\phi$, respectively.
Furthermore, the Hamilton-Jacobi equation is separable in oblate spheroidal coordinates, leading to a third nontrivial constant of motion, $\beta$, which is quadratic in momenta \cite{Landau}. 
By substituting  $\beta$ by $-Q-L_z^2-2a^2E$, the quantity $Q$ can be considered as the Newtonian analogue of Kerr's  Carter constant \cite{Carter} obtaining either one of the following forms \cite{Eleni}:
\begin{align}
   Q&=(1-\eta^2)p_{\eta}^2+\eta^2
\left(-2 E a^2+\frac{L_z^2}{1-\eta^2} \right)\label{Qh}\\
&=-p_{\xi}^2(\xi^2+1)+2a^2 E \xi^2+2 G M a\xi-\frac{L_z^2
\xi^2}{\xi^2+1}\label{Qj}. 
\end{align}

The existence of a third integral of motion  renders the Euler problem completely integrable in terms of quadratures; as there are three independent and in involution (i.e., $\{H_0,L_z\}=\{H_0,Q\}=\{L_z,Q\}=0$) integrals of motion as the number of the degrees of freedom of the Euler Hamiltonian system.
The expressions (\ref{Qh}) and (\ref{Qj}) are quite similar to the corresponding expressions relating the Carter constant in Kerr with either
$p_{\theta}$ and $\theta$, or $p_r$ and $r$.

An extensive list of key similarities that the  Euler potential shares with the gravitational field of the relativistic Kerr black hole can be found in Ref. \cite{Eleni}. The analogy between the two problems is better revealed by replacing $a \xi$ and $\eta$ of the Euler field with  $r$ and $\cos\theta$, respectively,  
mimicking the Boyer-Lindquist coordinates of Kerr.
Actually, the equations of motion in a Kerr metric at 1st Post-Newtonian order and at large $r$-values reduces to
the equations of motion in the Euler field \cite{Mama}

 
\section{The perturbed Euler}

We perturb the Euler field in order to find a Newtonian analogue of a slightly perturbed Kerr spacetime, by adding
a small point mass $m$ ($m<<M$) at the origin of the axes.
In this case the expression of the quadrupole moment and all higher mass moments are different from those of the unperturbed Euler, now obeying the following relation:
\begin{align}
    M_0&=M+m, \label{M0}\\
    M_{2l}&=(-a^2)^{l} M,
\end{align}
with $l=1,2...$. The multipole moments $M_k$ with odd $k$ vanish due to the reflection symmetry along the equatorial plane.
The new potential in oblate spheroidal coordinates takes the form:
 \begin{equation}\label{V1}
     V(\xi,\eta)=-\frac{G M \xi}{a(\xi^2+\eta^2)}-\frac{G m}{a\sqrt{1+\xi^2-\eta^2}}.
 \end{equation}
We will rewrite the potential in such a way that the unperturbed and the perturbed fields correspond to the same total mass $M_0$, so that both fields will be comparable with respect to their asymptotic limit at infinity:
\begin{equation}\label{veper}
      V(\xi,\eta)=-\frac{G M_0 \xi}{a(\xi^2+\eta^2)}+\frac{G m}{a} \left( \frac{\xi}{\xi^2+\eta^2}
      -\frac{1}{\sqrt{1+\xi^2-\eta^2}} \right).
\end{equation}
Thus when $m=0$ the system degenerates into the integrable  Euler problem.

When an integrable Hamiltonian system becomes slightly perturbed, the new Hamiltonian can be written in terms of the old integrable Hamiltonian $H_0$ plus a perturbation term $H_1$:
\begin{equation}\label{pertHam}
    H=H_0+\epsilon H_1.
\end{equation}
In our case $H_0$ is the Hamiltonian given exactly
by Eq. (\ref{Ham}).
We assume that the mass $m$ is small enough, compared to $M_0$, to apply classical perturbation theory. 
The term $H_1$ is given by:
\begin{align}\label{Hamper}
     H_1=\frac{G M_0}{a} \left(\frac{\xi}{\xi^2+\eta^2}-\frac{1}{\sqrt{1+\xi^2-\eta^2}}\right),
\end{align}
while the perturbative parameter is defined by $\epsilon=m/M_0$.

The new Hamiltonian has no dependence either on the time variable $t$ or the azimuthal angle $\phi$, due to the stationarity and axisymmetry of the new potential. 
As a result, there are two constants of motion; the energy:
\begin{equation}
       E=H=\frac{a^2}{2} (\xi^2+\eta^2)\left[
      \frac{\Dot{\xi}^2}{1+\xi^2} + \frac{\Dot{\eta}^2}{1-\eta^2}\right] 
 +\frac{a^2}{2}(1+\xi^2)(1-\eta^2) \Dot{\phi}^2+V(\xi,\eta),\label{En}
\end{equation}
and the component of angular momentum along the axis of symmetry:
\begin{equation}
    L_z=p_{\phi}=a^2(\xi^2+1)(1-\eta^2)\dot{\phi}\label{Lz}.
\end{equation}
However, the Hamilton-Jacobi equation is not separable anymore; there is no third integral of motion, which is independent and in involution with the energy, $E$, and the $z-$angular momentum, $L_z$.
In the next sections we will numerically confirm that, by investigating the Poincar{\'e} maps of orbits in the potential (\ref{veper}) and find properties related to non-integrability, such as chaotic motion and Birkhoff chains, as long as $m \neq 0$. 

As long as we are interested in bound orbits, 
we could define an effective potential $V_{\rm eff}$ to 
rewrite \eqref{En} as
\begin{equation}\label{Veffeq}
     0=\frac{1}{2}a^2 (\xi^2 + \eta^2)
\left(\frac{\dot{\xi}^2}{\xi^2+1} +
\frac{\dot\eta^2}{1-\eta^2} \right)+V_{\rm eff}(\xi,\eta),
\end{equation}
with 
\begin{equation}\label{Veff}
     V_{\rm eff}(\xi, \eta)=
     \frac{L_z^2}{2 a^2(\xi^2+1)(1-\eta^2)}+V(\xi,\eta)-E,
\end{equation}
where \eqref{Lz} has been used to replace the centrifugal part of the kinetic energy.


From Eq. (\ref{Veffeq}), it is obvious that the motion is allowed only for $V_{\rm eff}\leq 0$.
When an orbit reaches the curve $V_{\rm eff}=0$, the velocities $\dot{\xi}$ and $\dot{\eta}$ become zero (turning points); thus the  curve $V_{\rm eff}=0$ is known as Curve of Zero Velocity (CZV) \cite{AposGeraCont}.
Bound orbits are allowed in the interior of a closed CZV where the effective potential is negative. 
Additionally  bound orbits are characterized by $E<0$,
since orbits with $E\geq 0$ have CZVs that are not closed but are extended to infinity.
The number and the size of the distinct allowed regions on the poloidal plane $(\xi,\eta)$, within which a bound orbit could evolve,
 depend on the values of $E$ and $L_z$ of the orbit itself.

On the equatorial plane, i.e. at $\eta=0$, the effective potential reads:
\begin{equation}
    V_{\rm eff,eq}=-E+\frac{L_z^2}{2a^2(\xi^2+1)}
    -\frac{GM_0}{a\xi}\left[1-\epsilon\left(1-\frac{\xi}{\sqrt{1+\xi^2}}\right)\right].
\end{equation}
Especially a circular equatorial orbit (CEO) 
at $\xi=\xi_0$ satisfies:
\begin{equation}
V_{\rm eff,eq}(\xi_0)=\left.\frac{\partial V_{\rm eff,eq}}{\partial \xi}\right|_{\xi_0}=0.
\end{equation}
Solving the system of the last two equations we obtain the constants of motion for a CEO:
\begin{align}
    L_z=&\pm \frac{G M_0(\xi_0^2+1)}{\xi_0^{5/2}}
    \left[1-\epsilon \left(
    1-\frac{\xi_0^3}{(1+\xi_0^2)^{3/2}}\right)\right]\\
    E=&-\frac{G M_0}{2a\xi_0^3} (\xi_0^2-1) \left[1-  \epsilon \left( 1-
    \frac{\xi_0^3}{(\xi_0^2-1)\sqrt{1+\xi_0^2}} \right) \right]
\end{align}
For stable circular equatorial orbits we should have  $\left.\frac{\partial^2 V_{\rm eff,eq}}{\partial \xi^2} 
\right|_{\xi_0}\geq 0$.
An innermost stable circular orbit
(ISCO) exists when $\left.\frac{\partial^2 V_{\rm eff,eq}}{\partial \xi^2} 
\right|_{\xi_{\rm ISCO}}=0$.  
The perturbed Euler, as well as the corresponding unperturbed one, has an ISCO with the corresponding value of $\xi_{\rm ISCO}$ depending only on the perturbative parameter $\epsilon$. For $\epsilon=0$, $\xi_{\rm ISCO}=\sqrt{3}$, see \cite{Eleni}.

\section{KAM tori and resonant tori}
Due to the integrability of the Euler problem, bound orbits lie on two-dimensional tori in the 4-dimensional phase-space ($\xi,\dot{\xi},\eta,\dot{\eta}$), characterized by the three  integrals of motion. 
Tori corresponding to orbits that are characterised by the same $E$ and $L_z$ but different $Q$  are nested within each other.
Using action-angle variables one can define the orbit's characteristic  frequencies \cite{Eleni} of libration type $(\Omega_{\xi},\Omega_{\eta})$ associated with $\xi$ and $\eta$ oscillations. 
If the ratio of frequencies   $\Omega_{\xi}/\Omega_{\eta}$ is an irrational number, the motion will never repeat itself and it will gradually cover the whole torus (quasi-periodic orbit).
When the ratio of frequencies is a rational number  (resonance),
instead, the orbit repeats itself after an integer number of windings on the corresponding resonant torus (periodic orbit).

The Poincar{\'e} surface of section is a two-dimensional surface that intersects transversely the foliage of tori
\cite{Licht}. 
In our case, we have chosen as a surface of section
the plane $(\xi,\dot{\xi})$, when the orbit pierces the equatorial plane, $\eta=0$, with positive $\dot{\eta}$.
The Poincar{\'e} surface of section of each torus forms an invariant closed curve, which is either covered densely
when the orbit is quasi-periodic, or consisting of 
finite fixed points when the orbit is periodic.

When an intergable Hamiltonian system becomes slightly perturbed, the Kolmogorov-Arnold-Moser (KAM) theorem \cite{KAM1,KAM2,KAM3} states that almost all tori (the non-resonant ones)  become slightly deformed.
Thus the quasi-periodic orbits survive under a sufficiently small perturbation. They are confined on a 2-torus (KAM torus) which is slightly deviating from the unperturbed one. Consequently   the corresponding surface of section resembles the surface of section of the initially integrable system, but with a slightly deformed shape; these are the invariant KAM curves of the perturbed system.

The resonant tori, instead, 
are destroyed, when the system is slightly perturbed, 
according to the Poincar{\'e}-Birkhoff theorem \cite{PoincBirk1,PoincBirk2},
 forming Birkhoff chains of islands on the Poincar{\'e} section.
These islands are built around the fixed points of the initial unperturbed (integrable) system.
The interior of these islands consists of a new family of
KAM curves all sharing the same rational ratio of fundamental frequencies as the corresponding resonant torus of the unperturbed system.

The Birkhoff islands of stability are very thin and their detection on a Poincar{\'e} section could be quite tedious.
A useful method  to study nonintegrable systems and numerically detect the location of a chain of islands is the so-called
rotation number, which  actually gives the ratio of the fundamental frequencies \cite{Contopoulos}.
The rotation number $ \nu_{\theta} $ is defined by:
\begin{equation}
     \nu_{\theta}=\lim_{N\to\infty}\frac{1}{2\pi N}\sum_{i=1}^{N}\theta_i,
\end{equation}
with $N$ denoting the number of crossings of the Poincar{\'e} section by a phase-space trajectory. 
The angles of rotation $\theta_i $ are calculated as follows: 
at first one finds on the surface of section the fixed central point ${\bf u}_0$, which corresponds to the spherical orbit, $\xi=const$, and around which all KAM curves of quasi-periodic orbits are formed.
The position 
${\bf R}_i = {\bf u}_i-{\bf u}_0 $ of each crossing point $ {\bf u}_i $ of a phase space trajectory  on the surface of section with respect to $ {\bf u}_0 $ is defined.
Finally the angles $\theta_i = angle ({\bf R}_ {i + 1}, {\bf R}_i) $ between two successive positions of ${\bf R}_{i}$ on
the surface of section are calculated.

The rotation number is intimately related to the ratio of fundamental frequencies of the orbit itself.
On KAM tori the ratio of frequencies is irrational and varies from one curve to the other, so the rotation number changes continuously and monotonically as a function of the distance from the center ${\bf u}_0$ of the Poincar{\'e} section.
Its monotonic evolution is interrupted though, creating a plateau, by the islands of the periodic orbits, where the value of $ \nu_{\theta} $  is rational and fixed for all these orbits belonging to the same chain of islands. All these orbits are characterized by the same frequency ratio, regardless of the specific KAM curve to which each orbit belongs; within an island of stability, the ratio of frequencies remains constant, even though the frequencies themselves change from one KAM curve to another.

\subsection{The Poincar{\'e} section of the perturbed Euler problem}

\begin{figure}[ht]
    \centering
    \begin{subfigure}[b]{0.49\textwidth}
          \centering
    \includegraphics[width=\textwidth]{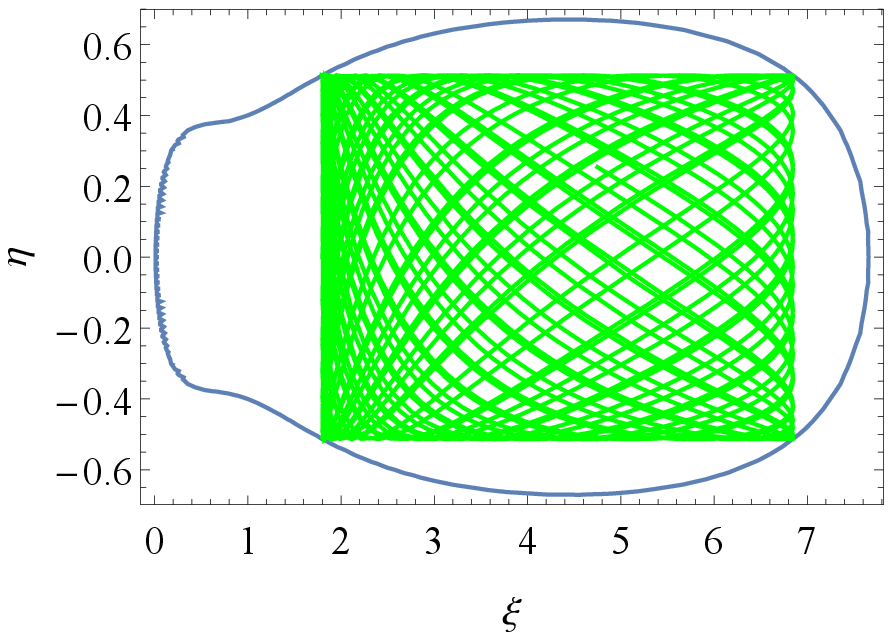}
    \caption{}
    \label{fig:1a}
          \end{subfigure}
      \hfill
  \begin{subfigure}[b]{0.49\textwidth}
          \centering
    \includegraphics[width=\textwidth]{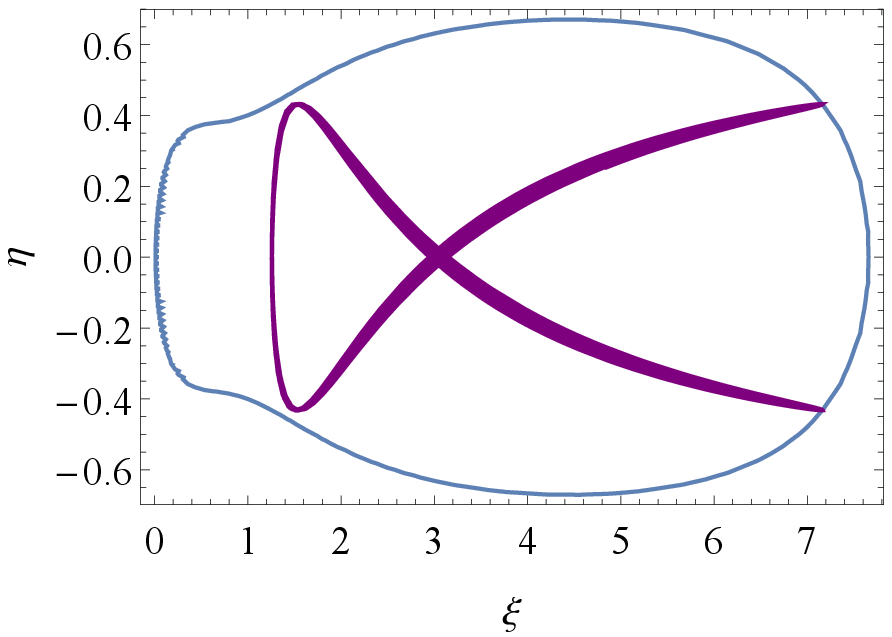}
    \caption{}
    \label{fig:1b}
        \end{subfigure}
    \caption{The CZV (blue boundary) of orbits in a perturbed  Euler field with $M_0=1, a=0.7, \epsilon=10^{-2}$. 
    The orbits are characterized  by orbital parameters $E=-0.156393$, $L_z=1.32878$.
   The left panel corresponds to an orbit with $\xi(0)=1.800$ (which leads to a KAM torus in phase space), while  the right panel corresponds to a fine-tuned orbit with $\xi(0)=1.257$ (which leads to a resonant KAM curve enclosed in a Birkhoff island on the Poincar{\'e} of section).
   Both orbits are evolved for the same total time: $T=500$.}
     \label{fig:1}
\end{figure}

In order to demonstrate that the perturbed Euler problem is non integrable we have constructed a Poincar{\'e} section and searched for Birkhoff islands. 
The physical parameters $M_0,a$, as well as the orbital parameters $E,L_z$ of the perturbed system with
$\epsilon=10^{-2}$ were initially chosen so that there are bound orbits. 
For such a fixed set of parameters, we evolved numerically a set of orbits with different initial conditions ($\xi(0),\dot{\xi}(0)=0,\eta(0)=0$, while
the initial velocity $\dot{\eta}(0)$ was calculated directly by Eqs. (\ref{En}, \ref{Lz}) apart from its sign, which was chosen to be positive); see Figs. \ref{fig:1a} and \ref{fig:1b}.

\begin{figure}[ht]
    \centering
    \begin{subfigure}[b]{0.49\textwidth}
          \centering
    \includegraphics[width=\textwidth]{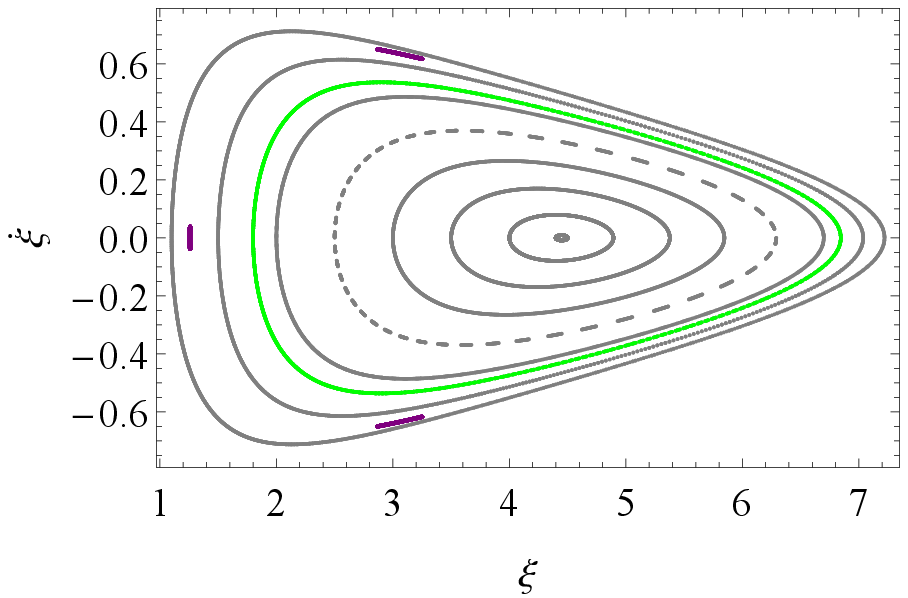}
    \caption{}
    \label{fig:2a}
          \end{subfigure}
      \hfill
  \begin{subfigure}[b]{0.49\textwidth}
          \centering
    \includegraphics[width=\textwidth]{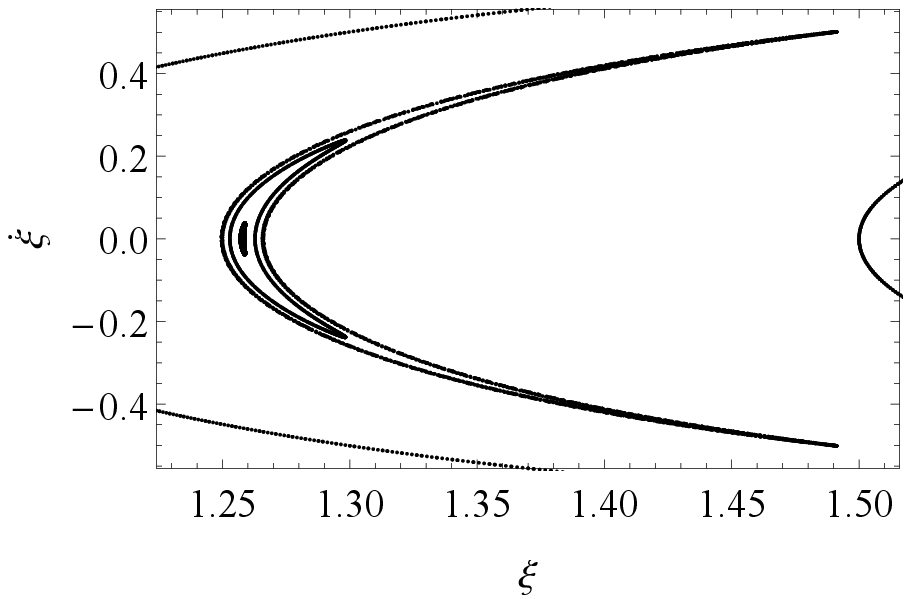}
    \caption{}
    \label{fig:2b}
        \end{subfigure}
    \caption{On the left panel, the Poincar{\'e} sections of 
   a number of
    orbits, all characterized by $E=-0.156393$, and $L_z=1.32878$
    (the same as for the previous Figure \ref{fig:1}), are drawn.
    Each orbit is evolved starting from a different initial condition $\xi(0)$. Most of the orbits lead to KAM curves (among them is the green KAM curve of the orbit shown in Fig. \ref{fig:1a}). Even the apparent dashed curve is a normal KAM curve that needs longer evolution time to fill the whole invariant curve. Also shown is the (purple) chain of Birkhoff islands that correspond to an orbit with resonance $\Omega_{\xi}:\Omega_{\eta}=2:3$. This is exactly the orbit of Fig. \ref{fig:1b}. On the right panel
    a detail of the Poincar{\'e} section of Fig. \ref{fig:2a}
    is drawn around the purple leftmost island. A few other Poincar{\'e} sections are shown, all corresponding to the same Birkhoff island of resonance $2:3$.}
     \label{fig:2}
\end{figure}

Then we formed the Poincar{\'e} section of all these orbits (see Fig. \ref{fig:2}) and measured the rotation number of each one. 
Most of them formed KAM-curves on the Poincar{\'e} section. 

By choosing the particular initial condition $\xi(0)$ that had led to three single fixed points on the Poincar{\'e} section  in the unperturbed problem corresponding to the resonance of $2:3$,
and assuming the same parameters $M_0,a,E,L_z$,
we managed to locate the chain of Birkhoff islands
of the corresponding non-integrable system.
Fiddling around this initial condition we
found a whole set of resonant orbits, 
belonging to the same chain of Birkhoff islands.

\begin{figure}[ht]
    \centering
    \begin{subfigure}[b]{0.49\textwidth}
          \centering
    \includegraphics[width=\textwidth]{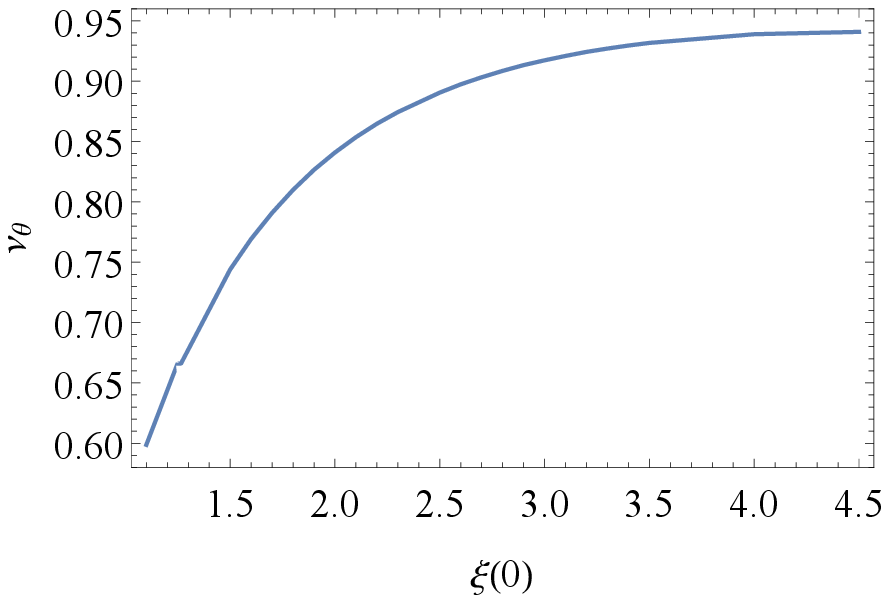}
    \caption{}
    \label{fig:3a}
          \end{subfigure}
      \hfill
  \begin{subfigure}[b]{0.49\textwidth}
          \centering
    \includegraphics[width=\textwidth]{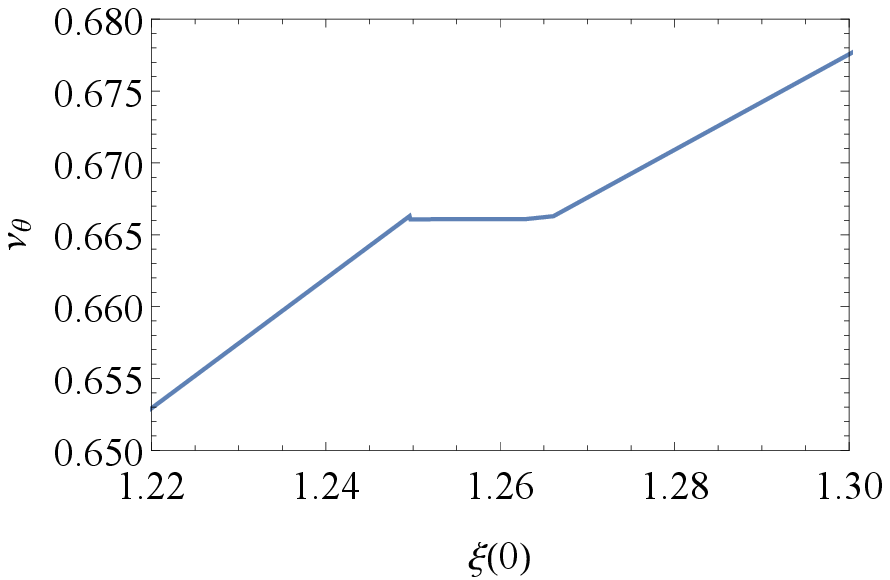}
    \caption{}
    \label{fig:3b}
        \end{subfigure}
    \caption{On the left plot, the rotation number $\nu_{\theta}$
    as a function of $\xi(0)$ is drawn for orbits with the same physical parameters as the ones presented in the two previous Figures. The horizontal axis spans almost
    the whole range of allowed $\xi(0)$'s up to the fixed central point ${\bf u_0}$. Apart of the anticipated monotonic character of $\nu_{\theta}(\xi(0))$, it is clear that around $\xi(0)=1.25$ there is narrow plateau corresponding to the particular resonance of $2:3$. A detail of this plateau is shown on the right panel. The tiny ``glitch'' on the left side of the plateau is an indication
    that the Birkhoff island is surrounded by a  very narrow chaotic 
    strip.}
     \label{fig:3}
\end{figure}

We have also drawn the rotation curve (Fig. \ref{fig:3}) 
of all orbits evolved. The strictly monotonic function
$\nu_{\theta}(\xi(0))$ is interrupted by a narrow plateau
corresponding to all orbits at resonance $2:3$; see Fig. \ref{fig:3b}. 

The width of the islands is intimately related to the magnitude of the perturbation $\epsilon$. More specifically, for sufficiently small perturbation parameter, the width should scale as $\sqrt{\epsilon}$ \cite{Arnol,GLG2}. We have confirmed this 
theoretical relation by measuring the width of the leftmost island
of resonance $2:3$ along the $\xi$-axis, for a few 
values of $\epsilon$ in the range  $10^{-5}$ to $10^{-2}$.



\begin{figure}[ht]
    \centering
    \includegraphics{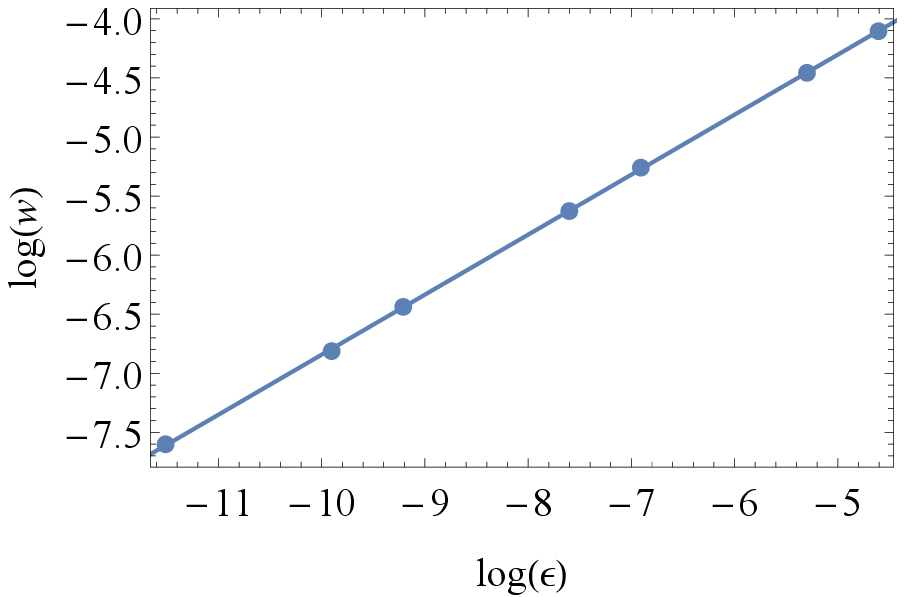}
    \caption{The width $w$ of the leftmost Birkhoff island of resonance $2:3$ has been computed for a few cases (shown as points) of  the perturbative parameter $\epsilon=m/M_0$. All points represent orbits with the same orbital parameters $E,L_z$ as in Fig. \ref{fig:1}. The best-fit straight line is the
        $\log(w)= -1.7641 + 0.507812 \log(\epsilon)$ which is in accordance with the expected theoretical slope of $1/2$
        (see \cite{GLG2}).}
    \label{width}
\end{figure}



\section{Inspirals}
\label{sec:5}

In the previous section we studied the evolution of orbits
in the perturbed Euler gravitational field alone, that is without
any other external force. This can be regarded as the analogue of the GR geodesic orbits in a specifically perturbed Kerr metric, like
the Manko-Novikov metric \cite{Manko}. The orbit of a compact object in a realistic EMRI though is not exactly geodesic, due to radiation reaction self-force. As long as the ratio of masses of the binary is sufficiently small, the orbits could be considered almost geodesics, but with adiabatically varying orbital parameters. This is true not only for EMRIs with a Kerr black hole as the central object, but with a non-Kerr supermassive central object as well.

In order to probe into the effect of resonance-crossing
due to an unknown self-force 
in a perturbed Kerr metric, we have used instead the 
perturbed Euler problem, endowed with an artificial dissipative ``self-force'', as a trustworthy toy model. Usually, the study of
such crossings in various perturbed Kerr background spacetimes is 
carried out by inducing the average value of energy and angular momentum losses to the corresponding geodesic equations of motion \cite{hybrid,AposGeraCont,destounis}.
Although this method leads in general to crude, though sufficiently accurate, adiabatic evolution of orbits,
when the orbit passes through a resonance, this approximation
becomes unreliable. The evolution of the orbit through a resonance under the instantaneous self-force itself,
could be quite different then. 

We have studied the evolution of orbits in
the perturbed Euler field, following two different schemes:
(i) By numerically integrating the second-order Euler-Lagrange equations of a test body under the specific Newtonian gravitational force, with a given external dissipative force, and (ii) by numerically integrating a new version of the equations of motion of the Newtonian field alone, suitably parametrized by the usual
integrals of motion, $E, L_z$, 
and imposing a prescribed time-dependence in $E, L_z$, caused by the 
same dissipative force. In Section \ref{sec:6} we will further explain
the new set of equations used under the second scheme.

The evolution of the first scheme describes, up to numerical errors,
the right evolution of the orbit, while the second scheme gives an approximate evolution of the orbit. When  the orbit is not at resonance, the two different schemes are expected to lead to approximately equivalent adiabatic evolutions at the limit of zero ``self-force''. Since the corresponding torus in phase-space is then densely covered, one should not anticipate any difference in the estimation of average losses, if these are measured either along 
a ``geodesic'' orbit (as in the second scheme), or along the actual orbit under the tiny ``self-force''.


In order to check how generic are our results with respect 
to the differences arising by using the two schemes described above, we have used two different dissipative forces, as analogues of the relativistic self-force. The general formula assumed for both external forces is
\begin{equation}\label{thisSF}
{\bf F}_{\rm ext}= - \delta  \mu f(\xi,\eta) {\bf v},
\end{equation}
where $\mu$ is the mass of the test particle,  $\bf v$ is its vector velocity on oblate spheroidal coordinates, $\delta<<1$ measures the magnitude of the ``self-force'', and the function $f(\xi,\eta)$ determines
how the strength of this force depends on the actual position
of the particle. The two cases investigated were
\begin{align}
f_1(\xi,\eta)&=1, \label{i1}\\
f_2(\xi,\eta)&=\frac{\sqrt{1-\eta^2}}{\xi}.  \label{i2}  
\end{align}
The first function $f_1$ corresponds to the usual atmospheric drag  force, while the second one, $f_2$ has been constructed so as to lead to a loss of energy and angular
momentum, while its strength is enhanced at 
lower $\xi$ values, where the field is stronger, 
and depends on the $\eta$-coordinate in a simple but 
physical, reflection-symmetric way.

The components of velocity ${\bf v}$ on spheroidal coordinates are (see Appendix A of \cite{Eleni}):
\begin{eqnarray}
v_{\xi}&=& a \dot{\xi}
\sqrt{\frac{\xi^2+ \eta^2}{\xi^2+1}},
\\
v_{\eta}&=& a \dot{\eta}
\sqrt{\frac{\xi^2+\eta^2}{1-\eta^2}},\\
v_{\phi}&=& a\dot{\phi}\sqrt{(1-\eta^2)(\xi^2+1)},
\end{eqnarray}

The instantaneous energy and $z$-angular momentum losses per unit mass are given from:
\begin{align}
    \left(\frac{dE}{dt}\right)_i 
    &= 
    {\bf v}\cdot {\bf a}_{\rm ext}
     &=&
     -\delta ~a^2 f_i(\xi,\eta) 
     \left[ (\xi^2+\eta^2)\left(\frac{\dot{\xi}^2}{1+\xi^2}+\frac{\dot{\eta}^2}{1-\eta^2}\right)+(1+\xi^2)(1-\eta^2)\dot{\phi}^2 \right], \\
    \left(\frac{dL_z}{dt}\right)_i 
    &=
    \hat{\bf z}\cdot ({\bf v}\times {\bf a}_{\rm ext}) 
    &=&-\delta ~ a^2 f_i(\xi,\eta)
    (1+\xi^2)(1-\eta^2)\dot{\phi},
\end{align}
 where    ${\bf a}_{\rm ext}={\bf F}_{\rm ext}/\mu$
 and $i$-index denotes the type of ``self-force'' used; see Eqs. (\ref{i1},\ref{i2}).
The averaged loss of either $E$ or $L_z$ at each 
orbital point is computed by 
\begin{equation}
    \left< \frac{d K}{dt} \right>
    =\lim_{T\to \infty}\frac{1}{T}\int_{0}^T \frac{d K}{dt} dt,
\end{equation}
where $K$ stands for $E$ or $L_z$, and the integrant
is computed along a ``geodesic'' orbit; i.e., an orbit where
no external force is applied, therefore $E, L_z$ remain constant.
The integration time $T$ need to be infinite so 
that the ``geodesic'' orbit has fully covered the whole available phase space for that orbit. Practically, we have integrated this ratio
for such a long time that the ratio converges to a finite value. Of course $T$
should be much longer than the scale of $\xi$ and $\eta$ oscillations.


\section{Orbital evolution from averaged energy and momentum losses.}
\label{sec:6}

In contrast to the Newtonian evolution of an orbit under a given instantaneous 
dissipative ``self-force'', which is straightforward in the case of
an orbit in pure or perturbed Euler potential,
the evolution due to the corresponding average losses 
of energy and $z$-angular momentum is quite more complicate.
The situation is exactly the opposite in the evolution
of an orbit in a perturbed Kerr metric; in that case the self-force itself
is not known (actually a complete analytic form is not known  for
a generic orbit, even in pure Kerr). However, one could easily evolve a geodesic orbit,
assuming the energy and the $z$-angular momentum are
given in analytic forms through a hybrid model \cite{hybrid} for
orbits in Kerr, suitably adjusted to accommodate for the
non-Kerr mass-quadrupole moment of the specific metric
\cite{GairLi}.

In order to evolve an orbit in a perturbed Euler potential,
with a given average loss of energy and $z$-angular momentum,
we cannot rely on Hamiltonian formalism,
since there is no straightforward way to turn a Hamiltonian
problem into a dissipative one, that its equations of motion lead to a given 
time-dependence of the integrals of motion of its non-dissipative counterpart.
We have overcome this issue by  transforming
the equations of motion
into a Hamiltonian-like form (that is to first order
differential equations), but suitably parametrized 
by quantities that are equivalent to
the integrals of motion when the ``self-force'' is absent. 

The new set of equations of motion
describing the orbit on the polar plane (the azimuthal angle
$\phi$ can be straightforwardly integrated once the angular momentum 
is given and the polar position is known as a function of time)
are differential equations for $\xi,\eta$ and an additional
angle $\theta$ defined as
\begin{align}
    \frac{\dot{\xi}}{\sqrt{1+\xi^2}}&=A \sin{\theta}\\
    \frac{\dot{\eta}}{\sqrt{1-\eta^2}}&=A \cos{\theta},
\end{align}
with $A$ being the positive definite kinetic energy along the polar plane
\begin{equation}
       A=
    \frac{\dot{\xi}^2}{\xi^2+1}+\frac{\dot{\eta}^2}{1-\eta^2}.
\end{equation}
The angle $\theta$ is a well defined quantity, related to the ratio of $\dot{\xi}$ and $\dot{\eta}$ terms in $A$, as long as $A$ is non-vanishing.


The new set of the equations of motion
 for $\xi,\eta,\theta$ (assuming the mass of the test particle is unity) then reads:
 \begin{align}
    \dot{\xi}&=A\sqrt{1+\xi^2}\sin{\theta},
    \label{1eqs}\\
    \dot{\eta}&=A\sqrt{1-\eta^2}\cos{\theta},
    \label{2eqs}\\
     \dot{\theta}&=-A\sin{\theta}\cos{\theta}H_1+\frac{A}{\xi^2+\eta^2}\times\nn\\
     &\biggl( H_2 H_3-2\sin{\theta}\cos{\theta}H_4+\frac{1}{A^2}\left[\frac{L_z^2H_5}{a^4}-\frac{G M_0}{a^3}\left((1-\epsilon)H_6+\epsilon H_7\right)\right]\biggr),\label{3eqs}
\end{align}
where
\begin{align}
    H_1&=\frac{\eta\sqrt{1+\xi^2}\cos{\theta}+\xi\sqrt{1-\eta^2}\sin{\theta}}{\sqrt{(1+\xi^2)(1-\eta^2)}}, \\
    H_2&=-(1-\eta^2)\sin^2{\theta}+(1+\xi^2)\cos^2{\theta},\\
    H_3&=\frac{\xi \cos{\theta}}{\sqrt{1+\xi^2}}+\frac{\eta \sin{\theta}}{\sqrt{1-\eta^2}},\\
    H_4&=\eta\sqrt{1-\eta^2}\cos{\theta}-\xi\sqrt{1+\xi^2}\sin{\theta},\\
    H_5&=\frac{\xi\sqrt{1-\eta^2}\cos{\theta}+\eta\sqrt{1+\xi^2}\sin{\theta}}{((1-\eta^2)(1+\xi^2))^{3/2}},\\
    H_6&=\frac{\sqrt{1+\xi^2}(\xi^2-\eta^2)\cos{\theta}-2\xi\eta\sqrt{1-\eta^2}\sin{\theta}}{(\xi^2+\eta^2)^2},\\
    H_7&=
    \frac{\xi\sqrt{1+\xi^2}\cos{\theta}+\eta\sqrt{1-\eta^2}\sin{\theta}}{(1+\xi^2-\eta^2)^{3/2}}.
\end{align}
The $A$ term in the set of equations above 
is simply a function of the total energy $E$ and $z$-angular momentum 
$L_z$, as well as of the coordinates $\xi,\eta$, through
\begin{equation}
    A = \sqrt{\frac{2}{a^2(\xi^2+\eta^2)} \left[
    E-\frac{L_z^2}{2 a^2(1+\xi^2)(1-\eta^2)}-
    V(\xi,\eta) \right]}~.
\end{equation}
The equation \eqref{3eqs} for $\theta$ has been derived by 
computing the time derivative of the ratio between 
the first two velocities (\ref{1eqs},\ref{2eqs}), in order to eliminate $A$,
and then introducing the expressions for $\ddot{\xi}$ 
and $\ddot{\eta}$ from the Euler-Lagrange equations
of the perturbed Euler field without any induced ``self-force'',
which are given in Appendix A.


Now Eqs. (\ref{1eqs},\ref{2eqs},\ref{3eqs}) form a set of three 
first-order  differential equations 
that describe the evolution of the system 
under the constrain of a constant energy and $z$-angular momentum.
As long as the $A$ term is non-vanishing, the evolution
is equivalent to that of Hamilton's equations.
However, if the $A$ term goes to zero, the set of the above equations
becomes indeterminate and one cannot use them to evolve the system.
The vanishing of $A$ term though correspond to a very special
set of initial conditions: when both $\dot{\xi}$ and $\dot{\eta}$
get simultaneously zero along the evolution. 
This situation arises when the orbit
touches the zero-velocity curve (CZV) which could be obtained by extremely fine tuning of initial conditions, corresponding to zero 
measure. 
Therefore we don't expect this
singular case to arise when arbitrary initial conditions
are evolved for a finite time. An orbit, though, might actually come very close to the CZV.
However,
one should not be worried about that, as long as the $A$ does not drop below a given threshold, leading to restricted numerical errors
in the evolution of the set of first-order differential equations
given above.

The advantage of
the new set of equations is that they give us the opportunity to
evolve the orbit with a predetermined time-varying 
law for $E$ and $L_z$. This is what we will 
exploit to compare the evolution of an orbit under
a ``self-force'' with the evolution of the orbit under the corresponding constant rate of change of energy and
$z$-angular momentum.

\begin{figure}
     \centering
     \begin{subfigure}[b]{0.49\textwidth}
         \centering
         \includegraphics[width=\textwidth]{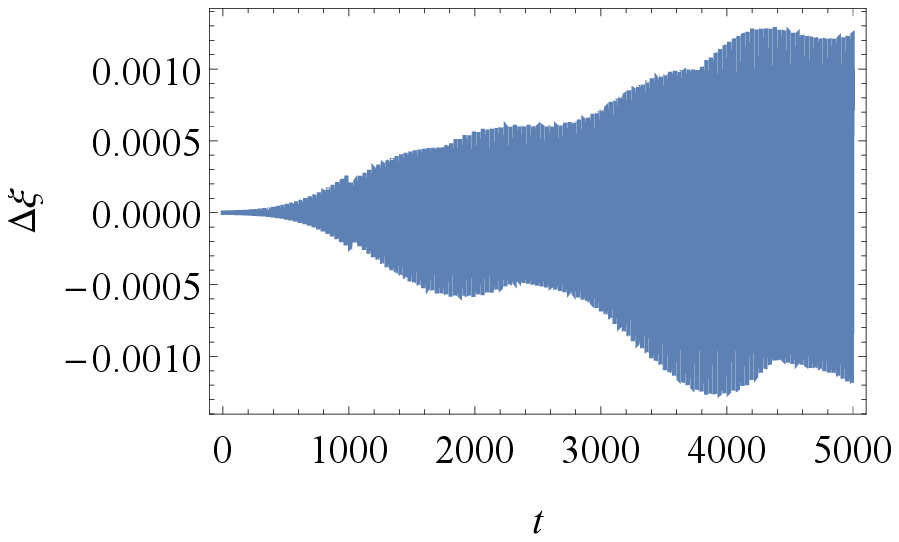}
         \caption{}
         \label{Aoft}
     \end{subfigure}
     \hfill
     \begin{subfigure}[b]{0.49\textwidth}
         \centering
         \includegraphics[width=\textwidth]{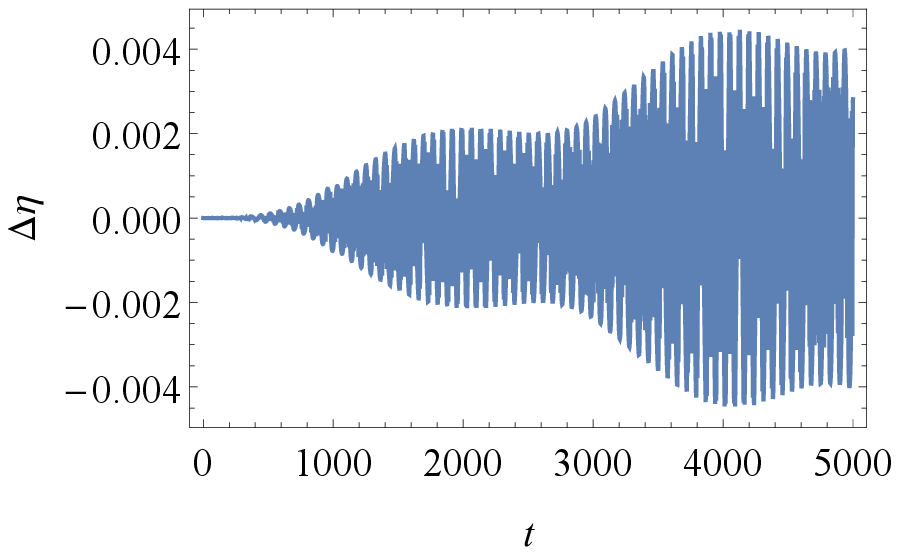}
         \caption{}
         \label{Eoft}
     \end{subfigure}
     \\
     \begin{subfigure}[b]{0.49\textwidth}
         \centering
         \includegraphics[width=\textwidth]{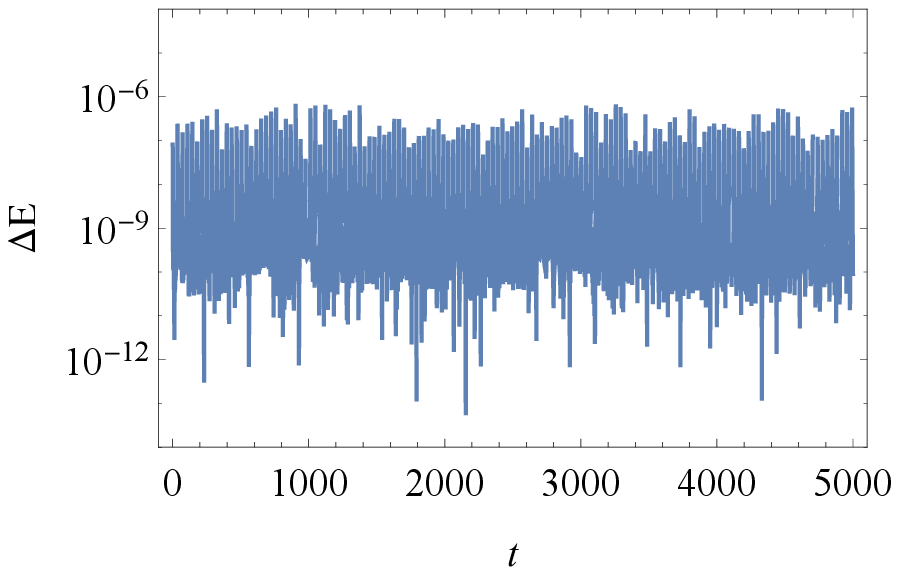}
         \caption{}
         \label{xioft}
     \end{subfigure}
     \hfill
     \begin{subfigure}[b]{0.49\textwidth}
         \centering
         \includegraphics[width=\textwidth]{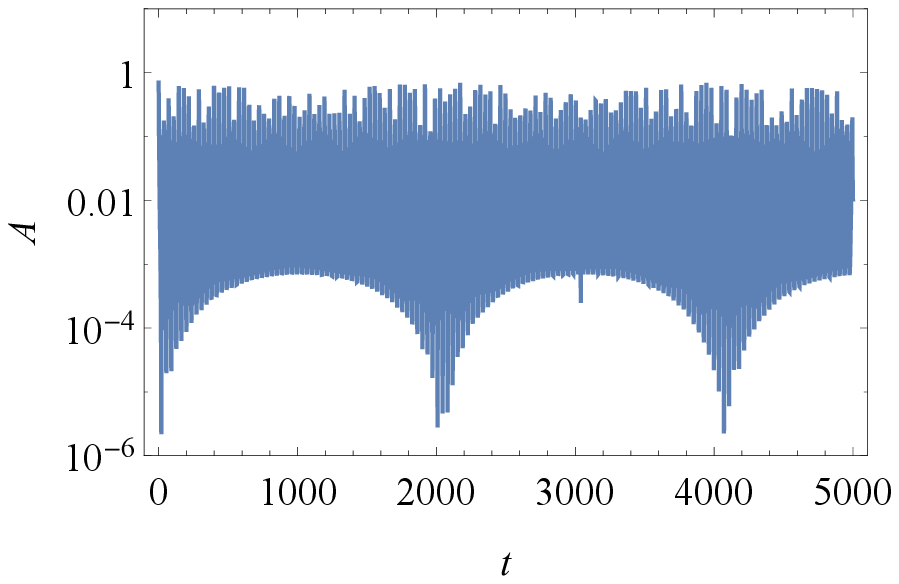}
         \caption{}
         \label{etaoft}
     \end{subfigure}
        \caption{The plots demonstrate the typical level of numerical accuracy in orbital evolution under the two integration schemes. In the upper two plots we have drawn the deviation of $\xi(t)$ and $\eta(t)$ between the two integration schemes. In the bottom-left plot (c) 
        the evolution of the deviation of the energy $E$, under the second scheme is presented. Finally, in the bottom-right plot (d)
        we
        have drawn the evolution of the parameter $A$ used in the second scheme (described in Sec. \ref{sec:6}). The orbital evolution in all these diagrams refers to an orbit with the same physical and orbital parameters used in Fig. \ref{fig:1}, and with
        initial conditions $\xi(0)=1.270,\eta(0)=\dot{\xi}(0)=0$.}
        \label{accuracy}
\end{figure}


%

\subsection{Accuracy Tests}

We have used MATHEMATICA in order to integrate numerically 
either scheme of orbital evolution. In order to test the numerical accuracy
of integration we first run the equations of motion \eqref{ELpert} for a few initial conditions without imposing any ``self-force''. We run also the system of equations (\ref{1eqs},\ref{2eqs},\ref{3eqs}) for the same initial conditions, 
with a constant value of the parameters $E,L_z$, equal to the energy and $z$-angular momentum corresponding to the initial conditions. Then we measure the orbital deviations between the two schemes. What we found (see Fig. \ref{accuracy})
is that there is some secular increase in the deviations of $\xi(t)$ 
and $\eta(t)$, caused by numerical errors, which are of order  
${\cal O}(10^{-3})$ for a total time of $5000$. As a comparison, the actual oscillations of $\xi$ and $\eta$ are of order $\sim 7$
and ${\cal O}(10^{-1})$, respectively,
with oscillation periods of order $\sim 30$ and $\sim 70$,
respectively. Moreover, we 
have tested the invariance of the conserved quantity $E$, under the
second scheme of integration. The deviations of $E$ did not exceed $10^{-6}$ for the same total time of evolution. 
Also, we have monitored the evolution of the parameter $A$,
along the integration, to ensure that 
the new set of equations does not lead to erroneous 
orbital evolution due to indeterminacy of the equations themselves.
In all cases we investigated the value of $A$ did not drop below $10^{-7}$
which is quite safe for the numerical accuracy of MATHEMATICA.


\section{Comparisons between the two schemes and Conclusions}
\label{sec:7}

The gravitational waves emitted by an EMRI,
the central source of which is not a pure 
Kerr black hole, are expected to demonstrate a peculiar
behavior when a resonance is met \cite{AposGeraCont, destounis}.
The ratio of the fundamental frequencies encoded in the signal will remain constant, while the system crosses a 
Birkhoff chain of islands. The duration of this crossing
is essential to discern such a non-integrable system,
describing the corresponding background spacetime.

In order to study the differences in crossing-times
of a given resonance arising from the evolution of the two
different schemes described in the Sec. \ref{sec:5},
we used a sequence of initial conditions quite close to the concave side of the leftmost Birkhoff island  of the $2:3$ resonance (see Fig. \ref{fig:2b}),
and evolved them directly with the instantaneous ``self-force'' scheme
up to the point where the particular resonance is hit. Subsequently,
we followed two different ways to further evolve the orbit: (i)
using the same scheme, up to the point where the orbit exits the corresponding Birkhoff island, and (ii) computing the average losses
of $E$ and $L_z$ at the specific phase-space coordinates when the 
orbit first enters the Birkhoff island and imposing these losses in 
Eqs. (\ref{1eqs},\ref{2eqs},\ref{3eqs}) of the second scheme until the orbit, again, exits the island. The $E$ and $L_z$ parameters introduced in these equations, through $A$ and $L_z$, are assumed to vary linearly with time,
with corresponding time-derivatives given by the losses mentioned above.

During the orbital inspiral, we periodically examined if the orbit is at resonance. 
This involved pausing the evolution using either integration scheme, then progressing the system along a ``geodesic'', assuming there were no ``self-force'',
and plotting its Poincar{\'e} section. The orbit
is at resonance, if a chain of Birkhoff islands forms on 
the Poincar{\'e} section.

\begin{figure}
     \centering
     \begin{subfigure}[b]{0.49\textwidth}
         \centering
         \includegraphics[width=\textwidth]{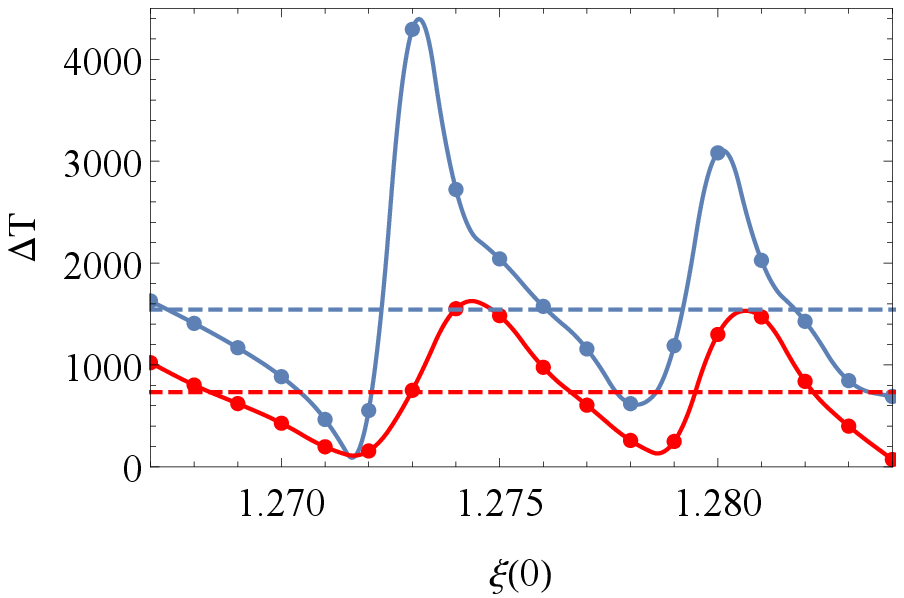}
         \caption{}
         \label{ratio1}
     \end{subfigure}
     \hfill
     \begin{subfigure}[b]{0.49\textwidth}
         \centering
         \includegraphics[width=\textwidth]{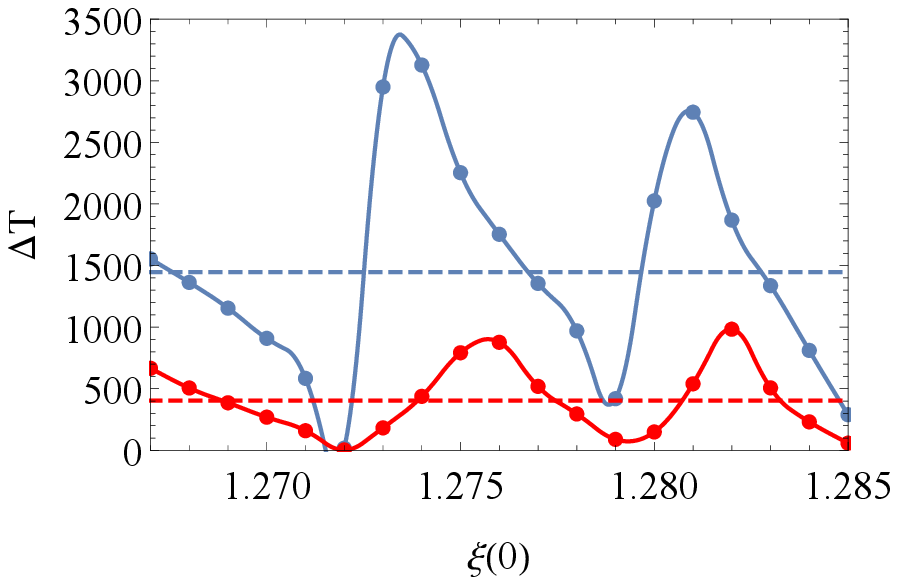}
         \caption{}
         \label{ratio2}
     \end{subfigure}
        \caption{These plots show how the crossing time for a Birkhoff island depends on the initial position of an orbit when (i) the orbit evolves under an instantaneous ``self-force'',(blue curves) or (ii) the orbit evolves by imposing a specific average loss of $E$, and $L_z$, which 
        corresponds to the particular ``self-force''. The left panel (a) is for the case of air-drag-type force ($f_1$) with $\delta=10^{-6}$, while the right panel is for the more complicated type of ``self-force'' ($f_2$) with $\delta=5 \times 10^{-6}$. The horizontal dashed lines represent the averages of all crossing-times in each case. The ups and downs of these plots are due to the fact that, depending on the entrance-point of the island, the orbital evolution through an island could be either very short or quite long.  }
        \label{crossings}
\end{figure}

For each unique evolution, we recorded the total time that the orbit spends within the island. The obtained results are presented in Fig. \ref{crossings}, illustrating the outcomes for the two types (\ref{i1},\ref{i2}) of ``self-force'' employed in our analysis. Depending on the entrance-point,
the evolution of an orbit inside a chain of Birkhoff islands
varies significantly: the orbit may get trapped at resonance for quite a long time, or pass the resonance in a very short period. This explains the recurrent ups and downs shown in the diagram, 
for both integration schemes in either type of ``self-force'' assumed.
This feature is reminiscent of the time intervals shown in Figure 11 of
\cite{AposGeraCont}, where the crossing time of the resonance $2:3$
for the  relativistic non-integrable case of Manko-Novikov was studied.

It is clear that the scheme based on average losses leads to
systematic and significant lower values of crossing times, 
compared to the crossing times under the instantaneous action of the ``self-force'' itself. The crossing time
due to  the actual evolution of the orbit is on average 2 to 3.5 
times larger than what one would get by imposing the constant values 
for $E$ and $L_z$ losses during the evolution.

Several distinct orbital evolutions were conducted using different types of ``self-force'', different magnitudes of $\delta$, and different orbital parameters $E,L_z$.
The crossing time, when the actual ``self-force'' was
employed to evolve the orbit, was boosted in all cases by a factor  similar, if not greater, to the case analyzed above.

The Newtonian analogue used in this paper is indicative of 
the differences arising in the evolution of an orbit
through a resonance of a slightly non-integrable system
under the two different integration schemes.
Moreover the similarity of the Kerr metric with the Euler problem 
suggests that these results are expected in a generically perturbed Kerr  
system.
Therefore, all estimations of the 
duration of the plateau effect in a slightly perturbed relativistic integrable 
system presented in the literature up to now
\cite{AposGeraCont,destounis}, might be suppressed,
compared to the actual duration of this effect in realistic EMRIs.

\section*{Acknowledgements}

Research was supported by the project of 
bilateral collaboration of scientists in Germany and Greece
IKYDA 2022.

\begin{appendices}

\section{Equations of motion}
The Lagrangian $L$ per unit mass of the perturbed Euler field is given by:
\begin{equation}
    L=p_{\mu}\dot{q}_{\mu}-H,
\end{equation}
where $H$ is the Hamiltonian of Eq. \eqref{pertHam}, $p_{\mu}$ are the conjugate momenta given in Eqs. (\ref{pxi},\ref{peta},\ref{pphi}) and $\dot{q}_{\mu}=(\dot{\xi},\dot{\eta},\dot{\phi})$. 
The equations of motion, that we solve numerically, are given by Euler-Lagrange equations:
\begin{align}\label{ELpert}
     \begin{split}
         \ddot{\xi}=&\frac{\xi}{\xi^2+\eta^2}\left(-\dot{\xi}^2\frac{1-\eta^2}{\xi^2+1}+\dot{\eta}^2\frac{\xi^2+1}{1-\eta^2}\right)-\frac{2\eta\dot{\eta}\dot{\xi}}{\xi^2+\eta^2}+\frac{\xi(\xi^2+1)(1-\eta^2)}{\xi^2+\eta^2}\dot{\phi}^2\\
         &-\frac{G(M_0-m)}{a^3}\frac{(\xi^2+1)(\xi^2-\eta^2)}{(\xi^2+\eta^2)^3}-\frac{Gm}{a^3}\frac{\xi(\xi^2+1)}{(\xi^2+\eta^2)(1+\xi^2-\eta^2)^{3/2}},\\
         \ddot{\eta}=&-\frac{\eta}{\xi^2+\eta^2}\left(-\dot{\xi}^2\frac{1-\eta^2}{\xi^2+1}+\dot{\eta}^2\frac{\xi^2+1}{1-\eta^2}\right)-\frac{2\xi\dot{\eta}\dot{\xi}}{\xi^2+\eta^2}-\frac{\eta(\xi^2+1)(1-\eta^2)}{\xi^2+\eta^2}\dot{\phi}^2\\
         &-\frac{G(M_0-m)}{a^3}\frac{2\xi\eta(1-\eta^2)}{(\xi^2+\eta^2)^3}-\frac{Gm}{a^3}\frac{\eta(1-\eta^2)}{(\xi^2+\eta^2)(1+\xi^2-\eta^2)^{3/2}},\\
         \ddot{\phi}=&\left(-\frac{2\xi\dot{\xi}}{\xi^2+1}+\frac{2\eta\dot{\eta}}{1-\eta^2}\right)\dot{\phi}.
     \end{split}
\end{align}
\end{appendices}
\printbibliography

\end{document}